\documentclass[12pt]{article}
\newcommand{\be}{\begin{equation}}
\newcommand{\ee}{\end{equation}}
\newcommand{\bea}{\begin{eqnarray}}
\newcommand{\eea}{\end{eqnarray}}
\newcommand{\sptwo}{1.4}
\newcommand{\doublespace}{\edef\baselinestretch{\sptwo}\Large\normalsize}
\newcommand{\newsection}[1]{\section{#1}\setcounter{equation}{0}}
\renewcommand{\theequation}{\thesection.\arabic{equation}}
\newcounter{newapp}
\renewcommand{\thenewapp}{\Alph{newapp}}
\begin{document}
\vspace*{0.2in}
\begin{center}
{\large\bf Non-BPS Brane Dynamics And Dual Tensor Gauge Theory}
\end{center}
\vspace{0.2in}
\begin{center}
T.E. Clark\footnote{e-mail address: clark@physics.purdue.edu}\\
{\it Department of Physics\\
Purdue University\\
West Lafayette, IN 47907-1396}\\
~\\
Muneto Nitta\footnote{e-mail address: nitta@th.phys.titech.ac.jp}\\
{\it Department of Physics\\
Tokyo Institute of Technology\\
Tokyo 152-8551, Japan}\\
~\\
T. ter Veldhuis\footnote{e-mail address: terveldhuis@macalester.edu}\\
{\it Department of Physics \& Astronomy\\
Macalester College\\
Saint Paul, MN 55105-1899}
\end{center}
\vspace{0.2in}
\begin{center}
{\bf Abstract}
\end{center}
The action for the long wavelength oscillations of a non-BPS p=3 brane embedded in N=1, D=5 superspace is determined by means of the coset method.  The D=4 world volume Nambu-Goldstone boson of broken translation invariance and the two D=4 world volume Weyl spinor Goldstinos of the completely broken supersymmetry describe the excitations of the brane into the broken space and superspace directions.  The resulting action is an invariant synthesis of the Akulov-Volkov and Nambu-Goto actions.  The D=4 antisymmetric tensor gauge theory action dual to the p=3 brane action is determined.
~\\
\pagebreak
\doublespace

\newsection{Introduction}

Brane world scenarios in which our four dimensional (D=4) world is assumed to be 
realized on a solitonic brane embedded in a higher dimensional space have been the topic of much research activity ~\cite{brane-world}.  Superstring theories, which may possibly provide the ultraviolet completion for such models, also require supersymmetry (SUSY) in higher dimensional space-time.  The minimum number of supercharges in D=5 or D=6 is eight, hence one is led to consider solitons in SUSY theories with at least eight supercharges.  BPS solitons preserve some fraction of the SUSY of the underlying model and so the soliton localized effective field theory also possesses it.  $\frac{1}{2}$-BPS domain walls breaking eight supersymmetries down to four were discussed in nonlinear sigma models as well as gauge theories (see ~\cite{8SUSY-sigma}-\cite{INOS1} and references therein).\footnote{Other types of solitons in these models were discussed in \cite{NNS}.}  $\frac{1}{2}$-BPS vortices in N=1, D=6 (eight supersymmetries) gauge theories were also discussed (see \cite{Eto:2004ii} and the references therein).  Since the effective field theory on BPS solitons 
is still supersymmetric, in the end, their SUSY must be broken if they are to realize our world.  This problem can be resolved if stable non-BPS branes exist which break SUSY completely on their world volume theory. 

A non-BPS domain wall in a generalized  N=1, D=4 (four supersymmetries) Wess-Zumino model 
was discovered in which its world volume is N=0, D=3 (no supersymmetries) \cite{Chibisov:1997rc}.  The long wavelength effective action for such a non-BPS brane was constructed using the method of nonlinear realizations
\cite{Volkov:jx}-\cite{Bagger:1996wp} in case the N=1, D=4 super-Poincar\'e symmetries are broken down to N=0, D=3 Poincar\'e symmetries \cite{Clark:2002bh} and equivalently by the Green-Schwarz method \cite{GS}, as generalized by Sen \cite{Sen:1999md} to the non-BPS case \cite{Clark:2004kz}.  Since the above model with a D=3 world volume is a toy model for the brane world picture, higher dimensional models are desired in order to realize a
non-BPS D=4 world-volume.  Some N=1, D=5 (eight supersymmetries) short distance models admitting non-BPS walls have been found recently~\cite{EMS} and the second reference in~\cite{INOS1}.  In the former \cite{EMS}, a periodic configuration of BPS and anti-BPS walls was considered in a N=1, D=5 $U(1)$ SUSY gauge theory with two hypermultiplets of equal charge.  There it was shown that the configuration was stable for small fluctuations and meta-stable for large fluctuations.  Of further interest is the second model in the second reference in~\cite{INOS1} in which the existence of non-BPS walls was shown, although no explicit solution has been constructed yet.  Briefly, this model consists of a D=5, N=1 SUSY $U(N_{\rm C})$ gauge theory with $N_{\rm F}$ hypermultiplets belonging to the fundamental representation.   If the Fayet-Iliopoulos term is added, the model contains $N_{\rm F}!/ (N_{\rm C}! (N_{\rm F} - N_{\rm C})!)$ discrete degenerate vacua~\cite{ANS}.  The general solution for $\frac{1}{2}$-BPS domain walls connecting these vacua was constructed~\cite{INOS1}.  However, there exists one pair of vacua which are connected by a non-BPS domain wall in the case of $N_{\rm F}=4$ and $N_{\rm C}=2$.  There exist more non-BPS walls connecting different sets of vacua for larger $N_{\rm F}$ and/or $N_{\rm C}$.  

The purpose of this paper is to construct the low energy thin domain wall effective action for the non-BPS case by means of the coset method.  In particular the nonlinear realization describing the breakdown of the N=1, D=5 super-Poincar\'e group to the (N=0) D=4 Poincar\'e group is given.  The appendix contains the N=1, D=5 SUSY algebra expressed in terms of the unbroken D=4 Lorentz group decomposition of the charges as a centrally extended N=2, D=4 SUSY algebra.  The low energy D=4 world volume fields consist of the Nambu-Goldstone boson scalar field $\phi$ corresponding to the broken D=5 space translation invariance and the two D=4 Weyl spinor Goldstino fields $\theta_\alpha$ and $\lambda_\alpha$, $\alpha =1,2$, of the broken N=1, D=5 supersymmetry.  

In section 2, the coset method is used in order to construct the effective action for the domain wall oscillations in the thin wall limit.  This p=3 brane action is an invariant synthesis of the Akulov-Volkov and Nambu-Goto actions with $\phi$ describing the space oscillations of the brane and $\theta$ and $\lambda$ describing the oscillations of the brane into the Grassmann directions of N=1, D=5 superspace.  The static gauge action has the form of the determinant of the induced metric vierbein $e_\mu^{~a}$
\be
\Gamma = -\sigma \int d^4 x \det{e} = -\sigma \int d^4 x \det{\hat{e}}\det{N} ,
\label{AVNG}
\ee
where $\sigma$ is the brane tension, $\hat{e}_\mu^{~a}$ is the Akulov-Volkov vierbein
\be
\hat{e}_\mu^{~a} = \delta_\mu^{~a} + i( \theta \stackrel{\leftrightarrow}{\partial}_\mu \sigma^a \bar\theta +
\lambda \stackrel{\leftrightarrow}{\partial}_\mu \sigma^a \bar\lambda )
\ee
and $N_a^{~b}$ is the Nambu-Goto vierbein.  After application of the \lq\lq inverse Higgs mechanism" \cite{Ivanov:1975zq}, $N_a^{~b} = \delta_a^{~b} +\frac{\hat\nabla_a \phi \hat\nabla^b \phi}{(\hat\nabla \phi)^2} \left(\sqrt{1- (\hat\nabla \phi)^2} -1\right),$ where the Nambu-Goldstone boson covariant derivative, $\hat\nabla_a \phi$, is defined as $\hat\nabla_a \phi = \hat{\cal D}_a (\phi +i(\theta\lambda - \bar\theta\bar\lambda))$, with the Akulov-Volkov partial covariant derivative, $\hat{\cal D}_a$, given by $\hat{\cal D}_a = \hat{e}_a^{-1\mu} \partial_\mu.$  The determinant of the Nambu-Goto vierbein yields 
\be
\det{N} = \sqrt{1- [\hat{\cal D}_a (\phi +i (\theta\lambda -\bar\theta\bar\lambda))]^2} .
\ee

In D=4 an antisymmetric tensor gauge theory can be used to equivalently describe the Nambu-Goldstone bosons of broken internal symmetries \cite{Ha}.  Bagger and Galperin showed that a tensor field may be regarded as a Nambu-Goldstone mode for broken translational symmetry in the case of partially broken supersymmetry on a BPS soliton \cite{Bagger:1997pi}.  In section 3 the tensor gauge theory action dual to the Akulov-Volkov-Nambu-Goto action, equation (\ref{AVNG}), is constructed.  This construction demonstrates the similar role played by the tensor gauge field in the non-BPS domain wall case as in the internal symmetry \cite{Ha} and BPS domain wall \cite{Bagger:1997pi} cases.  It is a supersymmetric generalization of the tensor gauge theory action dual to the p=3 bosonic brane action.  The bosonic brane embedded in D=5 space-time has the Nambu-Goto action
\be
\Gamma_{\rm NG} = -\sigma \int d^4 x \sqrt{1-\partial_\mu \phi \partial^\mu \phi}.
\ee
The dual tensor gauge theory action can be found by introducing the Lagrange multiplier \cite{Tseytlin:1996it} field strength $F_\mu$ so that $V_\mu = \partial_\mu \phi$ 
\be
\Gamma_{\rm NG}=-\sigma \int d^4 x \left[ \sqrt{1-V_\mu V^\mu} + F_\mu( V^\mu - \partial^\mu \phi)\right].
\ee
The Nambu-Goldstone boson field equation implies that $\partial_\mu F^\mu =0$ with tensor gauge field solution
$F_\mu = \epsilon_{\mu\nu\rho\sigma} \partial^\nu B^{\rho\sigma}$.  The vector field $V_\mu$ equation of motion is algebraic and so can be eliminated to yield the dual tensor gauge theory action
\be
\Gamma_{\rm NG}= -\sigma \int d^4 x \sqrt{ 1+ F^\mu F_\mu}.
\ee
The coset method action for the N=1, D=5 super-Poincar\'e symmetries spontaneously broken to D=4 Poincar\'e symmetries includes an auxiliary vector field for the broken D=5 Lorentz transformations.  Eliminating the Nambu-Goldstone scalar boson $\phi$ instead of the auxiliary vector field by means of its Euler-Lagarange equation (see Ivanov, et al. in \cite{Ivanov:1999fw}) leads directly, as equivalently the Lagrange multiplier method does, to the dual tensor gauge theory action
\be
\Gamma =-\sigma \int d^4 x \sqrt{-\det{(\hat{g}_{\mu\nu} + F_\mu F_\nu )}}, 
\ee
where $\hat{g}_{\mu\nu}= \hat{e}_\mu^{~a} \eta_{ab} \hat{e}_\nu^{~b}$ is the induced Akulov-Volkov metric.

\newsection{The Coset Method And Brane Dynamics}

The action for a non-BPS p=3 brane embedded in N=1, D=5 superspace can be constructed by means of the coset method for the case of the breakdown of the N=1, D=5 super-Poincar\'e group, denoted $G$, to the unbroken D=4 Poincar\'e and $R$ symmetry groups, denoted $H=ISO(1,3)\times R$.  The technique begins with the coset element $\Omega \in G/SO(1,3)\times R$
\be
\Omega = e^{ix^\mu P_\mu} e^{i\phi(x)Z} e^{i[\theta^\alpha (x) Q_\alpha +\bar\theta_{\dot\alpha} (x) \bar{Q}^{\dot\alpha} 
+ \lambda^\alpha (x) S_\alpha +\bar\lambda_{\dot\alpha}(x) \bar{S}^{\dot\alpha} ]} e^{i v^\mu (x) K_\mu} ,
\ee
where the $x^\mu$ denote the D=4 space-time coordinates parameterizing the 
world volume of the 3-brane in the static gauge, while the Nambu-Goldstone 
fields, denoted by $\phi (x) , \theta_\alpha (x), \bar\theta_{\dot\alpha} (x), \lambda_\alpha (x), \bar\lambda_{\dot\alpha} (x)$ and $v^\mu (x)$, describe the co-volume target space excitations of the brane.  Taken together, they 
act as coordinates of the coset manifold $G/H$.  Multiplication of the coset elements $\Omega$ by group elements $g \in G$ from the left results in transformations of the space-time coordinates and the Nambu-Goldstone fields according to the general structure
\be
g \Omega = \Omega^\prime h .
\ee
The transformed coset element yields the world volume coordinate transformations and the total variations of the fields 
\be
\Omega^\prime = e^{i x^{\prime \mu} P_\mu} e^{i\phi^\prime(x^\prime)Z} e^{i[\theta^{\prime\alpha} (x^\prime) Q_\alpha +\bar\theta^\prime_{\dot\alpha} (x^\prime) \bar{Q}^{\dot\alpha} 
+ \lambda^{\prime\alpha} (x^\prime) S_\alpha +\bar\lambda^\prime_{\dot\alpha}(x^\prime) \bar{S}^{\dot\alpha} ]} e^{i v^{\prime\mu} (x^\prime) K_\mu} ,
\label{Omega}
\ee
while $h$ is a field dependent element of $SO(1,3)\times R$. 

According to the coset construction method, the vierbein, the covariant 
derivatives of the Nambu-Goldstone fields and the spin connection can be 
obtained from the Maurer-Cartan one-form.  The Maurer-Cartan one-form can 
be determined by use of the Feynman formula for the variation of an 
exponential operator along with the Baker-Campbell-Hausdorff formula.  So doing, the 
Maurer-Cartan one-form is secured
\bea
\Omega^{-1} d\Omega &=& i\left[ \omega^{a} P_a + \omega_Z Z + \omega_{Q}^\alpha Q_\alpha 
+\bar\omega_{Q\dot\alpha} \bar{Q}^{\dot\alpha} + \omega_{S\alpha} S_\alpha 
+\bar\omega_{S\dot\alpha} \bar{S}^{\dot\alpha} \right.\cr
 & &\qquad\qquad\left. + \omega_{K}^a K_a + \omega_{M}^{\mu\nu} M_{\mu\nu} + \omega_R R \right],
\eea
where the individual world volume one-forms are found to be
\bea
\omega^{a} &=& \left( dx^b +i [\theta \sigma^b d\bar\theta - d\theta \sigma^b \bar\theta 
+\lambda \sigma^b d\bar\lambda - d\lambda \sigma^b \bar\lambda ]\right)\times \cr
 & &\qquad \times\left( \delta_{b}^{~a} +(\cosh{2\sqrt{v^2}} -1)\frac{v_b v^a}{v^2} \right)
-d\left( \phi + i[\theta\lambda - \bar\theta\bar\lambda]\right) \frac{\sinh{2\sqrt{v^2}}}{\sqrt{v^2}}v^a  \cr
\omega_Z &=& d\left(\phi + i[\theta\lambda - \bar\theta\bar\lambda] \right) \cosh{2\sqrt{v^2}} \cr
& &\qquad - \left( dx^a +i [\theta \sigma^a d\bar\theta - d\theta \sigma^a \bar\theta 
+\lambda \sigma^a d\bar\lambda - d\lambda \sigma^a \bar\lambda ]\right) v_a \frac{\sinh{2\sqrt{v^2}}}{\sqrt{v^2}} 
\cr
\omega_{Q}^{\alpha} &=& \cosh{\sqrt{v^2}} d\theta^{\alpha}
-\frac{\sinh{\sqrt{v^2}}}{\sqrt{v^2}}( d\bar\lambda \bar{\rlap{/}{v}})^{\alpha}\cr
\bar\omega_{\bar{Q}\dot\alpha} &=& \cosh{\sqrt{v^2}} d\bar\theta_{\dot\alpha}
+\frac{\sinh{\sqrt{v^2}}}{\sqrt{v^2}}( d\lambda \rlap{/}{v})_{\dot\alpha}\cr
\omega_{S}^{\alpha} &=& \cosh{\sqrt{v^2}} d\lambda^{\alpha}
+\frac{\sinh{\sqrt{v^2}}}{\sqrt{v^2}}( d\bar\theta \bar{\rlap{/}{v}})^{\alpha}\cr
\bar\omega_{\bar{S}\dot\alpha} &=& \cosh{\sqrt{v^2}} d\bar\lambda_{\dot\alpha}
-\frac{\sinh{\sqrt{v^2}}}{\sqrt{v^2}}( d\theta \rlap{/}{v})_{\dot\alpha}\cr
\omega_{K}^a &=& dv^a + \frac{iv^2}{2} \sinh{2\sqrt{v^2}} dv_b [\eta^{ba} -\frac{v^b v^a}{v^2}]\cr
\omega_{M}^{ab} &=& \left( \cosh{2\sqrt{v^2}} - 1 \right) \frac{(v^a dv^b -v^b dv^a)}{2v^2} \cr
\omega_R &=& 0 .
\label{oneforms}
\eea

The two sets of coordinate basis differentials $dx^\mu$ and $\omega^a$ are 
related to each other through the vierbein $e_\mu^{~a}$
\be
\omega^a = dx^\mu e_\mu^{~a} .
\ee
From equation (\ref{oneforms}) this yields, recalling $d=dx^\mu \partial_\mu$,
\bea
e_\mu^{~a} &=& \left(\delta_\mu^{~b} + i[\theta \sigma^b \stackrel{\leftrightarrow}{\partial}_\mu \bar\theta 
+ \lambda \sigma^b \stackrel{\leftrightarrow}{\partial}_\mu \bar\lambda ]\right) \left( 
\delta_b^{~a} + \left( \cosh{2\sqrt{v^2}} -1 \right) \frac{v_b 
v^a}{v^2}\right) \cr
& &\qquad - \frac{\sinh{2\sqrt{v^2}}}{\sqrt{v^2}}v^a \partial_\mu [\phi +i(\theta\lambda - \bar\theta\bar\lambda)].
\label{e2}
\eea
Under a $G$-transformation the vierbein transforms with one world index and one 
tangent space (structure group) index as
\be
e_\mu^{\prime ~a}= G_\mu^{-1 \nu} e_\nu^{~b} L_b^{~a} .
\label{eprime}
\ee
Using equation (\ref{Omega}), $G^{~\mu}_{\nu}=\partial x^{\prime \mu}/\partial x^\nu$ and $L_a^{~b}$ is the D=4 Lorentz transformation corresponding to $h$ and has determinant one, $\det{L} = 1$.  The leading term in the N=1, D=5 super-Poincar\'e invariant action is given by the \lq\lq cosmological constant\rq\rq term
\be
\Gamma = -\sigma \int d^4 x \det{e} ,
\ee
with $\sigma$ denoting the brane tension parameter.  The lagrangian is the constant 
brane tension integrated over the hyperarea of the brane.  The action is 
invariant due to equation (\ref{eprime}) and since $d^4 x^\prime =d^4 x \det{G}$ and $\det{L}=1$.

The fully covariant vierbein, $e_\mu^{~a}$, can be factorized into the product of the partially covariant
Akulov-Volkov vierbein $\hat{e}_\mu^{~a}$,
\be
\hat{e}_\mu^{~a} = \delta_\mu^{~a} +i \left(\theta \sigma^a \stackrel{\leftrightarrow}{\partial}_\mu \bar\theta
+ \lambda \sigma^a \stackrel{\leftrightarrow}{\partial}_\mu \bar\lambda \right),
\ee
and the Nambu-Goto vierbein $N_b^{~a}$,
\be
N_b^{~a} = \delta_b^{~a} + (\cosh{2\sqrt{v^2}} -1) \frac{v_b v^a}{v^2} -\frac{\sinh{2\sqrt{v^2}}}{\sqrt{v^2}}
v^a \hat{\cal D}_b [\phi +i(\theta\lambda -\bar\theta\bar\lambda)],
\ee
where the partial covariant Akulov-Volkov derivative is defined by $\hat{\cal D}_a = \hat{e}_a^{-1 \mu}\partial_\mu$,
\be
e_\mu^{~a} = \hat{e}_\mu^{~b}N_b^{~a} .
\ee
Thus, the invariant action involves the product of the Akulov-Volkov determinant and the determinant of the Nambu-Goto vierbein, as in equation (\ref{AVNG}).  The latter can be evaluated to yield
\be
\det{N} = \cosh{2\sqrt{v^2}}\left[ 1 -\frac{\tanh{2\sqrt{v^2}}}{\sqrt{v^2}} v^a \hat{\cal D}_a (\phi + i(\theta\lambda -\bar\theta\bar\lambda))\right] .
\label{NGvphi}
\ee

The vector field $v^a$ associated with the broken D=5 Lorentz transformations appears in the action without any derivatives.  Hence, it is an auxiliary field and can be eliminated by means of its equation of motion.  Equivalently, the Maurer-Cartan 
one-form associated with the broken translation generator $Z$ can be $G$-covariantly set to zero.  This leads to the elimination of $v^a$ via the \lq\lq inverse Higgs mechanism" \cite{Ivanov:1975zq}.  Expressing the $\omega_Z$ one-form in terms of the 
partially covariant one-form $\hat\omega^a \equiv dx^\mu \hat{e}_\mu^{~a} = dx^a + i (\theta \sigma^a \stackrel{\leftrightarrow}{d} \bar\theta - \lambda \sigma^a \stackrel{\leftrightarrow}{d} \bar\lambda )$ (hence $d=dx^\mu \partial_\mu = \hat\omega^a \hat{\cal D}_a $) gives
\be
\omega_Z = \hat\omega^a \cosh{2\sqrt{v^2}} \left[  \hat{\cal D}_a \left(\phi 
+ i(\theta\lambda -\bar\theta\bar\lambda)\right) -\frac{\tanh{2\sqrt{v^2}}}{\sqrt{v^2}} v_a \right] .
\ee
Setting this to zero results in the \lq\lq inverse Higgs mechanism\rq\rq
\be
v_a \frac{\tanh{2\sqrt{v^2}}}{\sqrt{v^2}} = \hat{\cal D}_a \left( \phi + i(\theta\lambda -\bar\theta\bar\lambda)\right).
\ee
Substituting this into the determinant of the Nambu-Goto vierbein yields the 
SUSY generalization of the Nambu-Goto lagrangian
\be
\det{N} = \frac{1}{\cosh{2\sqrt{v^2}}}
= \sqrt{1- \left({\hat{\cal D}}_a \left[\phi + i(\theta\lambda -\bar\theta\bar\lambda)\right]  
\right)^2} .
\ee
Hence the complete ($G$-invariant) N=1, D=5 super-Poincar\'e invariant Akulov-Volkov-Nambu-Goto action for a non-BPS p=3 brane embedded in N=1, D=5 superspace is given by
\bea
\Gamma &=& -\sigma \int d^4 x \det{e} = -\sigma \int d^4 x \det{\hat{e}}\det{N} \cr
 &=&-\sigma \int d^4 x \left\{ \det{\left[ \delta_\mu^{~a} + i( \theta \stackrel{\leftrightarrow}{\partial}_\mu \sigma^a \bar\theta +\lambda \stackrel{\leftrightarrow}{\partial}_\mu \sigma^a \bar\lambda )\right]} \times \right. \cr
& &\left. \qquad\qquad\qquad\qquad \times \sqrt{1- \left({\hat{\cal D}}_a \left[\phi + i(\theta\lambda -\bar\theta\bar\lambda)\right]  
\right)^2}\right\} .
\eea

\newsection{The Dual Tensor Gauge Theory}

Returning to action (\ref{AVNG}) and equation (\ref{NGvphi}) for the Nambu-Goto determinant with all fields independent, the $\phi$ equation of motion yields the Bianchi identity for the dual field strength vector $F^\mu$ (see Ivanov, et al. in \cite{Ivanov:1999fw})
\be
0=\frac{\delta \Gamma}{\delta \phi} = -\partial_\mu F^\mu ,
\ee
where
\be
F^\mu = (\det{\hat{e}}) (v^a \hat{e}^{-1 \mu}_a ) \frac{\sinh{2\sqrt{v^2}}}{\sqrt{v^2}}  .
\ee
Since the dual of $F^\mu$ is closed, $F^\mu$ can be (locally) expressed as
\be
F^\mu =\epsilon^{\mu\nu\rho\sigma} \partial_\nu B_{\rho\sigma},
\ee
where the 2-form $B_{\mu\nu}$ is the tensor gauge potential.  The lagrangian can be expressed as
\be
\det{e}= \det{\hat{e}}\det{N} = \det{\hat{e}} \cosh{2\sqrt{v^2}} - F^\mu 
\partial_\mu \left[ \phi + i(\theta\lambda -\bar\theta\bar\lambda)\right] .
\label{tensorlagrangian}
\ee
Exploiting the definition of $F^\mu$ further so that
\be
\frac{v^a v^b}{v^2} = \frac{(F^\mu \hat{e}_\mu^{~a})(F^\nu 
\hat{e}_\nu^{~b})}{(F\hat{e})^2}
\ee
results in
\be
\cosh{2\sqrt{v^2}} = \sqrt{\left( 1+ \frac{(F\hat{e})^2}{(\det{\hat{e}})^2}\right)}.
\label{tensorcosh}
\ee
Integrating equation (\ref{tensorlagrangian}) over the world volume after having substituted equation (\ref{tensorcosh}) and integrating by parts in order to set $\partial_\mu F^\mu =0$, the tensor gauge theory action dual to the non-BPS p=3 brane 
Nambu-Goto-Akulov-Volkov action is obtained
\be
\Gamma = -\sigma \int d^4 x \sqrt{- (\det{\hat{g}}) + F^\mu \hat{g}_{\mu\nu} F^\nu } ,
\ee
where the Akulov-Volkov metric is given by $\hat{g}_{\mu\nu} = 
\hat{e}_\mu^{~a}\eta_{ab}\hat{e}_\nu^{~b}$.

From the definition of the dual field strength $F^\mu$ and its expression in terms of the tensor gauge potential, it is a world volume vector density.  Hence, it is convenient to define the covariant field strength $F_\mu$ according to
\be
F_\mu \equiv \frac{1}{\det{\hat{e}}} \hat{g}_{\mu\nu} F^\nu .
\ee
The dual tensor gauge theory action can then be written as
\be
\Gamma = -\sigma \int d^4 x \sqrt{- \det{\left( \hat{g}_{\mu\nu} + 
F_{\mu}F_{\nu}\right)}} .
\ee

\section{Discussion}
In this paper the effective action describing the dynamics of the
Nambu-Goldstone degrees of freedom localized on a non-BPS p=3 brane
embedded in N=1, D=5 superspace was determined by means of the coset
method. It was shown that the Nambu-Goldstone field associated with the
broken translational symmetry can be described equivalently in terms of
either a scalar field or an anti-symmetric tensor gauge potential. Besides
the Nambu-Goldstone fields associated with the broken space-time
symmetries, in principle other massless degrees of freedom may appear in the
low energy effective action, depending on the details of the underlying
model. For example, additional massless scalar fields occur in
configurations of multiple non-interacting parallel domain walls, where
moduli correspond to distances between pairs of walls. These
non-Nambu-Goldstone massless modes likely appear as \lq\lq matter'' fields coupled to the
Nambu-Goldstone modes in terms of the coset construction of the effective
low energy action. In addition, if in a particular underlying model some
internal global symmetries are broken by a domain wall configuration,
corresponding Nambu-Goldstone modes appear in the effective action. In
fact, this latter situation is often realized by solitons in non-Abelian
gauge theories (see for instance \cite{Eto:2004ii,Ritz:2002fm}). In this case, each such
Nambu-Goldstone boson corresponds to a broken non-Abelian global internal
symmetry. The dual action, therefore, is expected to contain a non-Abelian tensor (as was discussed
by Freedman and Townsend in \cite{Ha}), in addition to the Abelian tensor associated with the broken translational symmetry. Other directions in which the coset construction
can be extended is to include more than the minimal number of supersymmetry
generators or to allow for a higher number of soliton codimensions. Such
generalizations would be of importance in order to construct the low energy
action of massless degrees of freedom in many interesting models. For
instance, a non-BPS vortex with codimension two was recently found in a
theory with eight supercharges \cite{Markov:2004mj}. Since the low energy
effective action of the model considered in this model contains
Nambu-Goldstone bosons associated with internal global non-Abelian
symmetries broken by the vortex, its dual action is anticipated to contain a non-Abelian tensor in addition to Abelian tensors, as discussed before.
\bigskip
~\\
\noindent
The work of MN was supported by the Japan Society for the Promotion of Science under the Post-Doctoral Research Program while that of TEC was supported in part by the U.S. Department of Energy under grant DE-FG02-91ER40681 (Task B).

\setcounter{newapp}{1}
\setcounter{equation}{0}
\renewcommand{\theequation}{\thenewapp.\arabic{equation}}

\section*{\large\bf Appendix: \, N=1, D=5 Super-Poincar\'e Algebra}

The D=5 Poincar\'e symmetry generators consist of the energy-momentum operator $P^M$ and the angular momentum operator $M^{MN}$, with $M, N =0, 1, 2, 3, 4$, which obey the algebra (the D=5 metric is taken to be $\eta^{MN} =(+,-,-,-,-)$)
\bea
\left[M^{MN}, M^{RS}\right] &=& -i\left(\eta^{MR} M^{NS} -\eta^{MS}M^{NR} + \eta^{NR} M^{MR} - \eta^{NR} M^{MS}\right) \cr
\left[M^{MN},P^L\right] &=& i\left(P^M \eta^{NL} - P^N \eta^{ML}\right)\qquad\qquad\qquad \left[P^M, P^N\right] = 0 .
\label{ISO(1,4)}
\eea
The N=1, D=5 super-Poincar\'e algebra has in addition the 4 component complex (Dirac) supersymmetry charges ${\cal Q}_a$ and $\bar{\cal Q}_a ={\cal Q}_b^\dagger \gamma^0_{ba}$ and the $R$-symmetry automorphism generator, $R$.  The non-vanishing commutators are
\bea
\{{\cal Q}_a,\bar{{\cal Q}}_b\} = +2\gamma_{ab}^M P_M \quad & &\quad \left[M^{MN},{\cal Q}_a\right] = - \frac{1}{2} \gamma_{ab}^{MN} {\cal Q}_b \cr
\left[R,{\cal Q}_a \right] = + {\cal Q}_a \qquad\qquad & & \quad
\left[R,\bar{\cal Q}_a \right] = - \bar{\cal Q}_a .
\label{D=5SUSY}
\eea

The non-BPS domain wall breaks the N=1, D=5 super-Poincar\'e symmetry group $G$ to the D=4 Poincar\'e symmetry and $R$ symmetry groups denoted $H=ISO(1,3)\times R$.  The N=1, D=5 super-Poincar\'e charges can be written in terms of their unbroken SO(1,3) Lorentz group content.  The unbroken symmetry group $H$ is generated by the charges $P^\mu$, with $\mu = 0, 1, 2, 3$, corresponding to translations in D=4 space-time of the world volume, $M^{\mu\nu}$, with $\mu , \nu = 0, 1, 2, 3$, corresponding to D=4 world volume Lorentz transformations and $R$ corresponding to chiral $R$ symmetry transformations.  The remaining charges generating elements of $G/H$ are the broken N=1, D=5 super-Poincar\'e charges.  $Z=P_4$ generates translations in the broken fifth dimension and acts as the central charge in the equivalent extended N=2, D=4 SUSY algebra, $K^\mu = 2 M^{4\mu}$ generates the broken D=5 Lorentz transformations.  The 8 broken N=1, D=5 (N=2, D=4) supersymmetry generators are complex 4 component (Dirac) spinors, ${\cal Q}_a$ and $\bar{\cal Q}_a$.  The Dirac spinors can be written in the Weyl representation in terms of the 2 component N=2, D=4 complex Weyl spinor charges, $Q_\alpha$, $\bar{Q}_{\dot\alpha}$, $S_\alpha$ and $\bar{S}_{\dot\alpha}$, as
\be
{\cal Q}_a = \pmatrix{Q_\alpha \cr
i\bar{S}^{\dot\alpha}} \qquad  \qquad 
\bar{\cal Q}_a = \pmatrix{-iS^\alpha & \bar{Q}_{\dot\alpha}}
\ee
(The D=5 Dirac matrices are given in terms of the D=4 Dirac matrices, in the Weyl representation, $\gamma^M=(\gamma^\mu , i\gamma_5)$.)

In terms of these operators the N=1, D=5 super-Poincar\'e algebra of equations (\ref{ISO(1,4)}) and (\ref{D=5SUSY}) becomes the centrally extended N=2, D=4 SUSY algebra given by

$$
\left[M^{\mu\nu}, M^{\rho\sigma}\right] = -i(\eta^{\mu\rho}
             M^{\nu\sigma} - \eta^{\mu\sigma} M^{\nu\rho}
             + \eta^{\nu\sigma} M^\mu\rho -\eta^{\nu\rho} M^\mu\nu )
$$
$$\begin{array}{lll}
   \left[M^{\mu\nu}, P^\lambda\right] =
           i(P^\mu \eta^{\nu\lambda} - P^\nu \eta^{\mu\lambda}) & & 
   \left[M^{\mu\nu}, K^\lambda\right] =
            i(K^\mu \eta^{\nu\lambda} - K^\nu \eta^{\mu\lambda})
\end{array}
$$
$$\begin{array}{lll}
   \left[Z,K^\mu \right] = 2iP^\mu \qquad &
    \left[P^\mu, K^\nu\right] = 2i\eta^{\mu\nu} Z \qquad &
    \left[K^\mu, K^\nu\right] = 4iM^{\mu\nu}
\end{array}
$$
$$\begin{array}{c}
 \left\{ Q_{\alpha}, \bar{Q}_{\dot{\alpha}} \right\}
             = 2\sigma_{\alpha\dot{\alpha}}^\mu P_\mu =
     \left\{ S_\alpha, \bar{S}_{\dot{\alpha}}\right\}
\end{array}
$$
$$\begin{array}{lll}
   \left[ M^{\mu\nu}, Q_\alpha\right] = -\frac{1}{2}
         \sigma_\alpha^{\mu\nu\beta}  Q_\beta \qquad & &\qquad
      \left[ M^{\mu\nu}, \bar{Q}_{\dot{\alpha}}\right]
             = -\frac{1}{2} \bar{\sigma}_{\dot{\alpha}\dot{\beta}}
                   ^{\mu\nu} \bar{Q}^{\dot{\beta}} \cr
    \left[M^{\mu\nu}, S_\alpha\right] = - \frac{1}{2}
               \sigma_\alpha^{\mu\nu\beta} S_\beta \qquad & &\qquad
        \left[M^{\mu\nu}, \bar{S}_{\dot{\alpha}}\right]
               = - \frac{1}{2} \bar{\sigma}_{\dot{\alpha}\dot{\beta}}
            ^{\mu\nu} \bar{S}^{\dot{\beta}} \cr
    \left\{Q_\alpha, S_\beta\right\} = -2 \epsilon_{\alpha\beta} Z \qquad & &\qquad
    \left\{\bar{Q}_{\dot{\alpha}}, \bar{S}_{\dot{\beta}}\right\} =
            -2 \epsilon_{\dot{\alpha}\dot{\beta}}Z \cr
    \left[ K^\mu, Q_\alpha\right] = -i \sigma_{\alpha\dot{\alpha}}^\mu
       \bar{S}^{\dot{\alpha}}\qquad & &\qquad
     \left[K^\mu, \bar{Q}^{\dot{\alpha}}\right] =
          +i \bar{\sigma}^{\mu\dot{\alpha}\alpha}S_\alpha\cr
     \left[K^\mu S_\alpha\right] = 
        +i\sigma_{\alpha\dot{\alpha}}^\mu
      \bar{Q}^{\dot{\alpha}}\qquad & &\qquad
      \left[K^\mu, \bar{S}^{\dot{\alpha}}\right]
       = -i \bar{\sigma}^{\mu\dot{\alpha}\alpha}Q_\alpha\cr
       \left[R,Q_\alpha\right] =  
              + Q_\alpha \qquad & &\qquad 
     \left[R,\bar{Q}_{\dot{\alpha}}\right] = -\bar{Q}_{\dot{\alpha}}\cr
     \left[R, S_\alpha\right] = -S_\alpha  \qquad & &\qquad
     \left[R, \bar{S}_{\dot{\alpha}} \right]= + \bar{S}_{\dot{\alpha}}
   \end{array}
$$
\be
 ~
\ee

\end{document}